\documentclass[12pt]{iopart}
\usepackage{mathrsfs}
\usepackage{graphicx}
\usepackage{hyperref}
\usepackage{cite}

\usepackage{multirow}
\usepackage{mathrsfs}

\begin{document}

\title{Implementations of more general solid-state (SWAP)$^{1/m}$ and controlled-(swap)$^{1/m}$ gates }

\author{Wen-Qiang Liu and  Hai-Rui Wei*}

\address{School of Mathematics and Physics, University of Science and Technology Beijing, Beijing 100083, China}

\ead{*hrwei@ustb.edu.cn}

\begin{abstract}
Universal quantum gates are the core elements in quantum information processing. We design two schemes to realize more general (SWAP)$^{1/m}$ and controlled--(swap)$^{1/m}$ gates (for integer $m\geq1$)  by directing flying single photons to solid--state quantum dots.  The  parameter $m$ is easily controlled by adjusting two quarter--wave plates  and one half--wave plate. Additional computational qubits are not required to construct the two gates. Evaluations of the gates indicate that our proposals are feasible with current experimental technology.
\end{abstract}



\section{Introduction} \label{sec1}

Quantum information processing (QIP), which exploits the quantum features of superposition and entanglement, promises to outperform its classical counterparts in solving certain computationally demanding problems in terms of security or speed \cite{book}. Quantum computing has the potential to execute Shor's factorization algorithm \cite{Shor,Shor1} and Grover's search algorithm \cite{Grover,Grover1,Long}, which are dramatically faster than traditional algorithms. Quantum computation allows one to build more efficient, secure, and useful quantum computers than the existing classical ones \cite{book}. Quantum communication possesses unprecedented advantages over classical communication. It holds an arduous assignment for absolutely secure and reliable information dissemination as well as faithful transfer of unknown quantum states between distant sites.

It is well known that quantum gates are the core elements in QIP. Controlled-NOT (CNOT) gates are among the most popular universal gates and are essential for various quantum information protocols \cite{universal,optimal}.  (SWAP)$^{1/m}$ gates are the cheapest and the most natural two--qubit gates, and they can be used widely in quantum computation and communication  \cite{cheapest1}.  In 2008,  Balakrishnan and  Sankaranarayanan \cite{Balakrishnan} studied the entangling character of (SWAP)$^{1/m}$ gates.  It has been shown that the (SWAP)$^{1/m}$ family of gates is universal and as efficient as CNOT gates in terms of  the required gate count in performing arbitrary two--qubit quantum operations \cite{cheapest1}. Therefore,  (SWAP)$^{1/m}$ gates  provide a way to implement quantum computation independent of CNOT gates; three (SWAP)$^{1/m}$ gates and six single-qubit gates are sufficient and necessary to implement arbitrary two--qubit operations \cite{cheapest1}. Controlled-(swap) (Fredkin) gates, first introduced by Fredkin and Toffoli \cite{Fredkin},  play an important role in classical reversible computation and multi-user quantum communication \cite{book}. Controlled-(swap) gates and Hadamard gates are competent to synthesize any multiqubit unitary operation \cite{Fredkin}. Moreover, controlled--(swap) gates can be directly applied to error--correction \cite{error-correcting1,error-correcting3}, quantum fault tolerance \cite{fault-tolerrance}, quantum algorithms \cite{algorithm1}, fingerprinting \cite{fingerprinting,fingerprinting1}, optimal cloning \cite{optimal-cloning}, and controlled entanglement filtering.

Significant progress has been made on the (SWAP)$^{1/m}$ family of gates in recent years. Early in 1995, Barenco \emph{et al.} \cite{universal} proved that three CNOT gates are sufficient to implement a SWAP gate.  In 2005, Fiorentino \emph{et al.} \cite{Fiorentino} presented an interesting scheme to realize a linear--optical SWAP gate for the momentum and polarization degrees of freedom (DOFs) of a single photon. In the same year, Liang and Li \cite{Liang-Li} fulfilled the construction of a SWAP gate between a single photon and an atom. A few values of $m$ and the corresponding entangling character of (SWAP)$^{1/m}$  gates were proposed by Balakrishnan \emph{et al.} \cite{Balakrishnan} in 2008. Subsequently, in 2010, Koshino \emph{et al.} \cite{Koshino} showed  a photon--photon (SWAP)$^{1/2}$ gate via a three--level $\Lambda$ system. In 2015, Wei \emph{et al.} \cite{wei} designed a quantum circuit to implement a solid--state (SWAP)$^{1/m}$  gate assisted by diamond nitrogen--vacancy (NV) centers. In 2018, a passive swap operation was demonstrated based on a single photon and an atom in a material $\Lambda$-system \cite{passiveSWAP}, which provided a versatile building block to achieve universal quantum gates such as (SWAP)$^{1/2}$ \cite{swap/2}. In 2019, Calafell \emph{et al.} created a (SWAP)$^{1/2}$ gate with high fidelity and success rate by employing strong two--plasmon absorption in graphene nanoribbons \cite{swap/22019}.

The optimal cost of a controlled--(swap) gate is five two-qubit entangling gates  \cite{Smolin,Yu}. In 2006, Fiur\'{a}\v{s}ek \cite{Fiurasek1} designed a linear optical controlled--(swap) gate with $4.1\times 10^{-3}$ success probability.  Subsequently, utilizing partial--SWAP gates, Fiur\'{a}\v{s}ek \cite{Fiurasek} presented the linear optical controlled--(swap) gate again. In 2008, Gong \emph{et al.} \cite{Gong} further improved the success probability of the controlled--(swap) gate to 1/64 based on time entanglement and linear optics. In 2016, Wei \emph{et al.} \cite{wei-Fredkin} proposed a scheme of achieving  a two--photon four--qubit controlled-(swap) gate via NV defect centers. In 2017, Ono \emph{et al.} \cite{Ono} experimentally demonstrated a photonic controlled--(swap) gate. Over the same period, controlled--(swap) gates were also obtained in a hybrid system comprising flying photons and atomic ensembles \cite{QD-Fredkin}. In 2018, Ren \emph{et al.} \cite{ren-Fredkin} constructed a near--deterministic polarization--spatial hyperparallel controlled--(swap) gate with high fidelity and efficiency using  NV centers in the optical cavity.

Thus far, various apparatuses have been proposed to implement QIP, and solid--state devices emerged as one of the most popular candidates among physical systems due to their good scalability, possible coherent control, and readout of the single spins.  Photon--mediated interactions between solid--state matter (solid state as the quantum node and a photon as the quantum bus) are of fundamental importance in long--distance solid--state QIP \cite{solid-state1}. Single--photon Hadamard gates \cite{photonHadamard},  deterministic two--photon controlled--phase gates \cite{controlled-phase}, and single--photon transports \cite{Jung-Tsung-Shen} were designed through atom--mediated interactions.  Semiconducting quantum dots (QDs) in cavities are an outstanding service for solid-state qubits, and remarkable progress has been achieved in QD--spin QIP. QDs could be assembled experimentally in large arrays and designed to have certain characteristics. The electronic spin confined in a charged QD supports $\mu$s coherence time \cite{coherence100,coherence} and ps or fs time--scale single--qubit manipulations \cite{Hadamard1,Hadamard2,rotation3,rotation4}.  The manipulation and measurement of the QD spin can be achieved electrically \cite{electrically} or optically \cite{optically}. Researchers have devoted much effort to the integration of charged QDs with cavities to improve photon--QD interactions. Charged QDs amalgamated with nanophotonic micropillar cavities have been experimentally demonstrated recently \cite{integrate1,integrate2,double}.

In this paper, we propose two schemes to definitively implement solid--state (SWAP)$^{1/m}$ and controlled--(swap)$^{1/m}$ gates for single QD qubits embedded in double--sided microcavities. The electron spins in single charged QDs serve as gate qubits, whereas flying photons act as mediated matter (bus) for bridging distant QDs. The spin-dependent phase shifts on uncoupled photons are employed to construct (SWAP)$^{1/m}$ and controlled--(swap)$^{1/m}$ gates. Our schemes have the following characters: (1) They are generalized and integer is $m\geq1$; (2) the controllable parameter $m$ can be achieved by employing two quarter--wave plates (QWPs) and one half--wave plate (HWP); (3) additional electrons are not required in the construction of the two gates; and (4) our schemes are simpler than their synthesis--based and cross--Kerr--based counterparts.


\section{Solid-state (SWAP)$^{1/m}$ gate} \label{sec2}

\begin{figure}[tpb]  \label{level}
\begin{center}
\includegraphics[width=7 cm,angle=0]{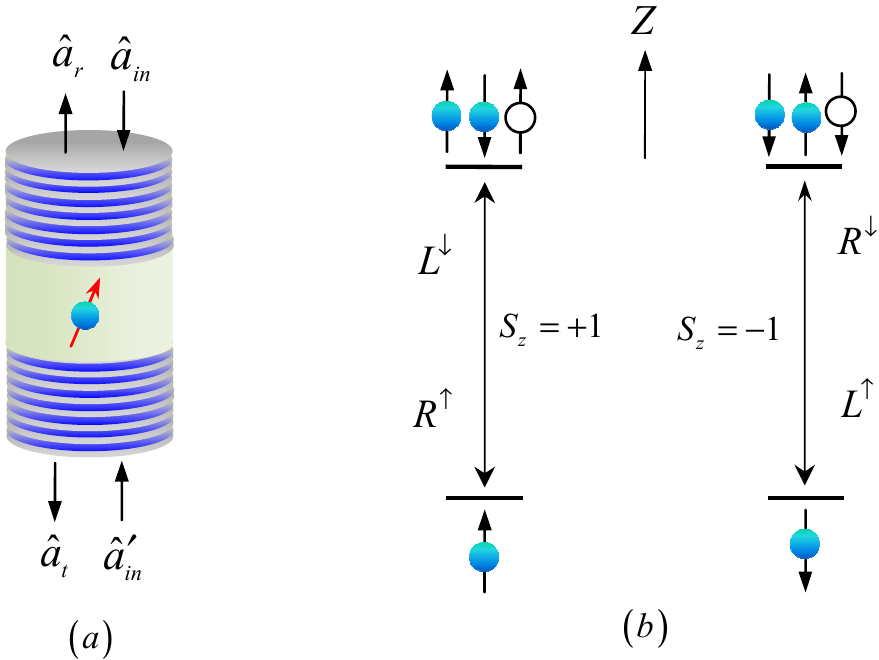}
\caption{(Color online) (a) Structure of a singly charged QD inside a double--sided optical microcavity with a circular cross--section. (b) Energy--level scheme of a singly charged QD inside a double--sided optical microcavity with the polarization allowed transition rules for the coupling photons. $\vert R\rangle$ ($\vert L\rangle$) represents a right--circularly (left--circularly) polarized photon.}
\end{center}
\end{figure}

Figure 1 illustrates the structure and energy level diagram for a negatively charged QD  trapped in a doubled--sided microcavity \cite{Hu2}. An excess electron is injected, and a negatively charged exciton $X^-$ consisting of two electrons of opposite spin  bounded to one heavy hole with two possible spin orientations is created by optical excitation \cite{trion-X}. A qubit is encoded in two ground electron spin states $|\uparrow\rangle$ and $|\downarrow\rangle$, with angular momentum projections $J_z=\pm1/2$. The exciton spin states are $|\uparrow\downarrow\Uparrow\rangle$ and $|\downarrow\uparrow\Downarrow\rangle$. Here, $|\Uparrow\rangle$ and $|\Downarrow\rangle$ are the heavy hole spin states with $J_z=\pm3/2$. Because of the conservation of total spin angular momentum, the transition $|\uparrow\rangle\rightarrow|\uparrow\downarrow\Uparrow\rangle$ is the resonance derived from the $S_z=+1$ circularly polarized photon, marked by $R^\uparrow$ and $L^\downarrow$. $|\downarrow\rangle\rightarrow|\downarrow\uparrow\Downarrow\rangle$ is the resonance originating from the $S_z=-1$ circularly polarized photon, marked by $R^\downarrow$ and $L^\uparrow$. Here, $R$ and $L$ denote the right-- and left--circularly polarized photons, respectively. The superscripts $\uparrow$ and $\downarrow$ of $R$ ($L$) indicate that the propagation directions of the $R$-- ($L$--) polarized  photons are along and against the $z$ axis, respectively.

The transmission and reflection coefficients of the incident photon are obtained by solving the Heisenberg equations of motion for cavity mode, which is driven by the corresponding input field operators $\hat{a}_{in}$  and $\hat{a}_{in}'$ from the top and bottom of the cavity, and from the dipole--lowering operations (i.e., $\hat{a}$ and $\sigma_-$) \cite{Heisenberg}
\begin{equation}          \label{eq1}
\cases{
\frac{d\hat{a}}{dt}=-\left[i(\omega_c-\omega)+\kappa+\frac{\kappa_s}{2}\right]\hat{a}-g\sigma_{-}-\sqrt{\kappa}\,\hat{a}_{in} - \sqrt{\kappa}\,\hat{a}_{in}'+\hat{H},\nonumber \\
\frac{d\sigma_-}{dt}=-\left[i(\omega_{X^-}-\omega)+\frac{\gamma}{2}\right]\sigma_{-}-g\sigma_z\hat{a}+\hat{G},}
\end{equation}
and the input--output relationships in the cavity are expressed as \cite{Heisenberg}
\begin{equation}        \label{eq2}
\hat{a}_r=\hat{a}_{in}+ \sqrt{\kappa}\,\hat{a},\qquad\qquad
\hat{a}_t=\hat{a}_{in}'+\sqrt{\kappa}\,\hat{a}.
\end{equation}
We take $\langle \sigma_z\rangle=-1$; that is, a sufficiently large $\kappa$ is taken by the external electrical field to ensure a weak excitation approximation  throughout our operation, which is applicable to  single--photon process, and even two--photon process \cite{applicable}. Thus, the reflection/transmission coefficients of a realistic QD--cavity system can be written as follows \cite{coefficients}
\begin{eqnarray}         \label{eq3}
r(\omega)=1+t(\omega),\quad
t(\omega)=\frac{-\kappa[i(\omega_{X^{-}}-\omega)+\frac{\gamma}{2}]}
           {[i(\omega_{X^{-}}-\omega)+\frac{\gamma}{2}][i(\omega_c-\omega)+\kappa+\frac{\kappa_s}{2}]+g^2}.
\end{eqnarray}
%
Here, $\omega_c$, $\omega$, and $\omega_{X^-}$ are the resonance frequencies of the cavity mode, the input incident single-photon pulse, and the dipole transition of the negatively charged exciton $X^-$, respectively.
$g$ is the corresponding QD--cavity coupling strength.
$\gamma/2$, $\kappa$, and $\kappa_s/2$ are defined as the decay rates of the $X^-$ dipole, the total cavity energy, and the cavity energy leakage, respectively.
$\hat{H}$ and $\hat{G}$ denote the noise operators related to reservoirs and are needed to conserve the commutation relations.
$\sigma_z$ is the inversion operator of the singly charged QD, and $\hat{a}_{in}$, $ \hat{a}'_{in}$ and $\hat{a}_{r}$, $\hat{a}_t$ are the input and output field operators, respectively.

When $\kappa\gg\kappa_s$ (side leakage from the cavity is negligible), under resonance frequency condition  $\omega_c=\omega=\omega_{X^-}$, the spin--dependent optical transitions can be written as \cite{transition1,transition2}
\begin{eqnarray}       \label{eq4}
\fl\qquad\quad|R^\uparrow\uparrow\rangle \rightarrow |L^\downarrow\uparrow\rangle,\;\;\;\;
|R^\downarrow\uparrow\rangle \rightarrow -|R^\downarrow\uparrow\rangle, \;
|R^\uparrow\downarrow\rangle \rightarrow -|R^\uparrow\downarrow\rangle, \;
|R^\downarrow\downarrow\rangle \rightarrow |L^\uparrow\downarrow\rangle,\nonumber\\
\fl\qquad\quad|L^\uparrow\uparrow\rangle \rightarrow -|L^\uparrow\uparrow\rangle, \;\;
|L^\downarrow\uparrow\rangle \rightarrow |R^\uparrow\uparrow\rangle, \;\;\;\;
|L^\uparrow\downarrow\rangle \rightarrow |R^\downarrow\downarrow\rangle,\;\;\;\;
|L^\downarrow\downarrow\rangle \rightarrow -|L^\downarrow\downarrow\rangle.
\end{eqnarray}
The underlying mechanism is that, for coupled photon--trion interactions, the circularly polarized photon is reflected without any phase shift, whereas for decoupled photon--trion interactions, the photon is transmitted with a $\pi$-phase shift.

\begin{figure}[!h]  \label{SWAP}
\begin{center}
\includegraphics[width=7.5 cm,angle=0]{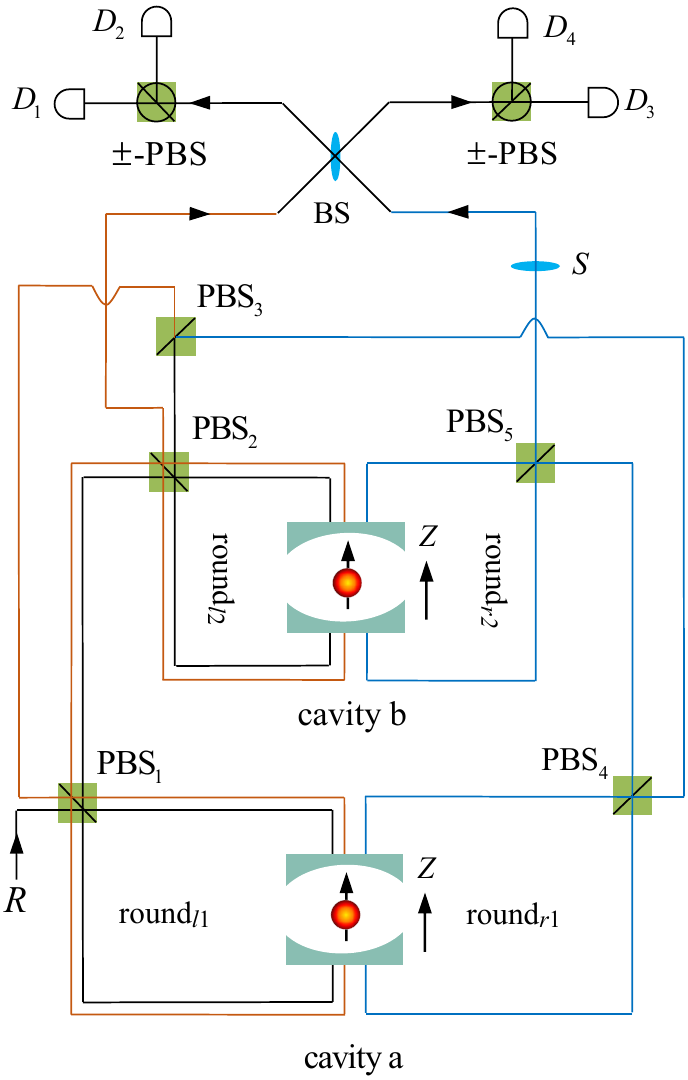}
\caption{(Color online) Schematic diagram for the construction of a deterministic solid-state (SWAP)$^{1/m}$ gate for $m\geq1$.  Each PBS represents a polarizing beam splitter in the $\{|R\rangle,\;|L\rangle\}$ basis transmitting the $R$--polarized component and reflecting the $L$--polarized component, respectively. $\pm$PBS transmits the $|+\rangle$--polarized photon and reflects the $|-\rangle$-polarized photon, respectively. $S$ is a phase gate performing the operation $S=e^{i\pi/m}|R\rangle\langle R|+|L\rangle\langle L|$ on the right arm. BS stands for the balanced beam splitter. Each $D$ denotes a single--photon polarization detector.}
\end{center}
\end{figure}

Next, we utilize above spin--dependent quantum phase to construct the (SWAP)$^{1/m}$  gate between two  independent QD spins, see figure \ref{SWAP}. The effect of such (SWAP)$^{1/m}$ gate can be presented by the unitary matrix
\begin{eqnarray}         \label{eq6}
\quad\quad\quad U_{{\rm(SWAP)}^{1/m}}=\left(\begin{array}{cccc}
1 & 0 & 0 & 0 \\
0 & \frac{1+e^{i\pi/m}}{2} & \frac{1-e^{i\pi/m}}{2} & 0 \\
0 & \frac{1-e^{i\pi/m}}{2} & \frac{1+e^{i\pi/m}}{2} & 0 \\
0 & 0 & 0 & 1
\end{array}\right),
\end{eqnarray}
in the computational  $\{|\uparrow_a\uparrow_b\rangle,\; |\uparrow_a\downarrow_b\rangle,\; |\downarrow_a\uparrow_b\rangle,\; |\downarrow_a\downarrow_b\rangle\}$ basis.

Firstly, suppose the two QDs are initially prepared into arbitrary normalization state
\begin{eqnarray}         \label{eq7}
|\Phi_e\rangle=(\cos\alpha|\uparrow_a\rangle +\sin\alpha|\downarrow_a\rangle)\otimes(\cos\beta|\uparrow_b\rangle+\sin\beta|\downarrow_b\rangle).
\end{eqnarray}

For the gate operation, a single photon in state $|R_l\rangle$ is injected. The polarizing beam splitter (PBS) in the $\{\vert R\rangle,\; \vert L\rangle\}$ basis  transmits the $R$-polarized  component to interact with the QD$_a$. That is, PBS$_1$ transforms the state of the whole system from
\begin{eqnarray}         \label{eq8}
|\Phi_{ep}\rangle_0=|R_l\rangle\otimes(\cos\alpha|\uparrow_a\rangle +\sin\alpha|\downarrow_a\rangle)\otimes(\cos\beta|\uparrow_b\rangle+\sin\beta|\downarrow_b\rangle)
\end{eqnarray}
into
\begin{eqnarray}         \label{eq9}
|\Phi_{ep}\rangle_1=|R_l^\downarrow\rangle\otimes(\cos\alpha|\uparrow_a\rangle +\sin\alpha|\downarrow_a\rangle)\otimes(\cos\beta|\uparrow_b\rangle+\sin\beta|\downarrow_b\rangle).
\end{eqnarray}
Here, subscripts $e$ and $ep$ stand for the electron state and the hybrid electron-photon state, respectively.
The interaction between $R$--polarized wave packet and QD$a$ induces the system from $|\Phi_{ep}\rangle_1$ to
\begin{eqnarray}         \label{eq10}
|\Phi_{ep}\rangle_2= (-\cos\alpha|R_l^\downarrow\rangle|\uparrow_a\rangle +\sin\alpha|L_l^\uparrow\rangle|\downarrow_a\rangle)\otimes(\cos\beta|\uparrow_b\rangle+\sin\beta|\downarrow_b\rangle).
\end{eqnarray}
Here, the subscripts $l$ or $r$ (mentioned later) of $R^\downarrow$ ($L^\uparrow$) denote that the $R^\downarrow$-- ($L^\uparrow$--) polarized component is emitted from the left or right arms, respectively (see figure \ref{SWAP}).

From equations (\ref{eq7})--(\ref{eq10}) combining with equation (\ref{eq4}), one can see that the transformation of round$_{l_1}$ in figure \ref{SWAP} can be described by a unitary matrix $U_{\rm round_{l_1}}$,
\begin{eqnarray}         \label{eq11}
\qquad\qquad\qquad U_{{\rm round}_{l_1}}=\left(\begin{array}{cccc}
-1 & 0 & 0  & 0 \\
0  & 0 & 0  & 1 \\
0  & 0 & -1 & 0 \\
0  & 1 & 0  & 0
\end{array}\right),
\end{eqnarray}
in the $\{|R\rangle|\uparrow\rangle,\; |R\rangle|\downarrow\rangle,\; |L\rangle|\uparrow\rangle,\; |L\rangle|\downarrow\rangle \}$ basis.

Subsequently, the photons interact with round$_{l2}$, and then the state of the system becomes
\begin{eqnarray}         \label{eq12}
|\Phi_{ep}\rangle_3&=
\cos\alpha\cos\beta|R_l^\downarrow\rangle|\uparrow_a\uparrow_b\rangle-
\cos\alpha\sin\beta|L_l^\uparrow\rangle|\uparrow_a\downarrow_b\rangle\nonumber\\&-
\sin\alpha\cos\beta|L_l^\uparrow\rangle|\downarrow_a\uparrow_b\rangle+
\sin\alpha\sin\beta|R_l^\downarrow\rangle|\downarrow_a\downarrow_b\rangle.
\end{eqnarray}

Next, as shown in figure \ref{SWAP}, the $L$--polarized components are reflected to interact with round$_{r1}$, round$_{r2}$, phase gate  $S=e^{i\pi/m}|R\rangle\langle R|+|L\rangle\langle L|$, and 50:50 balanced beam splitter (BS) in succession. Alternatively, the $R$--polarized components are transmitted to round$_{l1}$, round$_{l2}$, and  BS in succession.
Before and after the photons interact with round$_{r1}$ and round$_{r2}$ (round$_{l1}$ and round$_{l2}$),  Hadamard operations $H_{ea}$ and $H_{eb}$ are performed on QD$_a$ and QD$_{b}$, respectively. It is worth mentioning that the single--qubit operation performed on QD can be realized by using a ns spin resonance microwave pulse \cite{Hadamard1} or ultrafast ps (or fs) optical pulse from the cavity side \cite{Hadamard2} within the spin coherence time (20 $\mu$s) \cite{coherence}. When $(\pi/2)_y$ pulses are applied to complete $H_{ea}$ and $H_{eb}$, the incident photon can be stored in optical circular or  matter qubit.
$S=e^{i\pi/m}|R\rangle\langle R|+|L\rangle\langle L|$ can also be actualized by employing P$(\frac{2m+1}{2m}\pi)$, QWP$(\frac{\pi}{4})$, HWP$(\frac{m+1}{4m}\pi)$, QWP$(\frac{\pi}{4})$.
The transformations of phase shifter P$(\alpha)$, half--wave plate HWP$(\beta)$ and quarter--wave plate QWP$(\gamma)$ rotated by $\alpha$, $\beta$, and $\gamma$, respectively  can be written as \cite{QWP-HWP1}
\begin{eqnarray}         \label{eq6}
\qquad\qquad\qquad\quad U_{\rm{P}(\alpha)}=\left(\begin{array}{cc}
e^{i\alpha} & 0  \\
0 & e^{i\alpha}
\end{array}\right),
\end{eqnarray}
\begin{eqnarray}         \label{eq6}
\qquad\qquad\quad U_{\rm{HWP}(\beta)}=e^{i\frac{\pi}{2}}\left(\begin{array}{cc}
\cos 2\beta & \sin 2\beta  \\
\sin 2\beta & -\cos 2\beta
\end{array}\right),
\end{eqnarray}
\begin{eqnarray}         \label{eq6}
\qquad\qquad U_{\rm{QWP}(\gamma)}=\frac{1}{\sqrt{2}}\left(\begin{array}{cc}
1+i\cos 2\gamma & i\sin 2\gamma  \\
i\sin 2\gamma & 1-i\cos 2\gamma
\end{array}\right).
\end{eqnarray}
The operations of $H_{ea}$ and $H_{eb}$ can be described as
\begin{eqnarray}                     \label{eq13}
|\uparrow\rangle\leftrightarrow\frac{1}{\sqrt2}(|\uparrow\rangle+|\downarrow\rangle),\qquad\qquad
|\downarrow\rangle\leftrightarrow\frac{1}{\sqrt2}(|\uparrow\rangle-|\downarrow\rangle).
\end{eqnarray}
The transformations of 50:50 BS can be written as
\begin{eqnarray}                     \label{eq14}
|R_l\rangle\leftrightarrow\frac{1}{\sqrt2}(|R_l\rangle+|R_r\rangle),\qquad\quad\;  |L_l\rangle\leftrightarrow\frac{1}{\sqrt2}(|L_l\rangle+|L_r\rangle),\nonumber\\
|R_r\rangle\leftrightarrow\frac{1}{\sqrt2}(|R_l\rangle-|R_r\rangle),\qquad\quad  |L_r\rangle\leftrightarrow\frac{1}{\sqrt2}(|L_l\rangle-|L_r\rangle).
\end{eqnarray}
Therefore,  $H_{ea}$ and $H_{eb}$ change $|\Phi_{ep}\rangle_3$ to be
\begin{eqnarray}         \label{eq15}
\fl\qquad\qquad|\Phi_{ep}\rangle_4=
\frac{1}{2}\cos\alpha\cos\beta|R_l^\downarrow\rangle(|\uparrow_a\uparrow_b\rangle+|\uparrow_a\downarrow_b\rangle+|\downarrow_a\uparrow_b\rangle+|\downarrow_a\downarrow_b\rangle)\nonumber\\\quad-
\frac{1}{2}\cos\alpha\sin\beta|L_r^\uparrow\rangle(|\uparrow_a\uparrow_b\rangle-|\uparrow_a\downarrow_b\rangle+|\downarrow_a\uparrow_b\rangle-|\downarrow_a\downarrow_b\rangle)\nonumber\\\quad-
\frac{1}{2}\sin\alpha\cos\beta|L_r^\uparrow\rangle(|\uparrow_a\uparrow_b\rangle+|\uparrow_a\downarrow_b\rangle-|\downarrow_a\uparrow_b\rangle-|\downarrow_a\downarrow_b\rangle)\nonumber\\\quad+
\frac{1}{2}\sin\alpha\sin\beta|R_l^\downarrow\rangle(|\uparrow_a\uparrow_b\rangle-|\uparrow_a\downarrow_b\rangle-|\downarrow_a\uparrow_b\rangle+|\downarrow_a\downarrow_b\rangle).
\end{eqnarray}
Round$_{l1}$, round$_{l2}$, round$_{r1}$, and round$_{r2}$ transform $|\Phi_{ep}\rangle_4$ into
\begin{eqnarray}         \label{eq16}
\fl\qquad|\Phi_{ep}\rangle_5=
\frac{1}{2}\cos\alpha\cos\beta(|R_l^\downarrow\rangle|\uparrow_a\uparrow_b\rangle-|L_l^\uparrow\rangle|\uparrow_a\downarrow_b\rangle-|L_l^\uparrow\rangle|\downarrow_a\uparrow_b\rangle+|R_l^\downarrow\rangle|\downarrow_a\downarrow_b\rangle)\nonumber\\\fl\qquad\qquad\;\;\;-
\frac{1}{2}\cos\alpha\sin\beta(|L_r^\uparrow\rangle|\uparrow_a\uparrow_b\rangle+|R_r^\downarrow\rangle|\uparrow_a\downarrow_b\rangle-|R_r^\downarrow\rangle|\downarrow_a\uparrow_b\rangle-|L_r^\uparrow\rangle|\downarrow_a\downarrow_b\rangle)\nonumber\\\fl\qquad\qquad\;\;\;-
\frac{1}{2}\sin\alpha\cos\beta(|L_r^\uparrow\rangle|\uparrow_a\uparrow_b\rangle-|R_r^\downarrow\rangle|\uparrow_a\downarrow_b\rangle+|R_r^\downarrow\rangle|\downarrow_a\uparrow_b\rangle-|L_r^\uparrow\rangle|\downarrow_a\downarrow_b\rangle)\nonumber\\\fl\qquad\qquad\;\;\;+
\frac{1}{2}\sin\alpha\sin\beta(|R_l^\downarrow\rangle|\uparrow_a\uparrow_b\rangle+|L_l^\uparrow\rangle|\uparrow_a\downarrow_b\rangle+|L_l^\uparrow\rangle|\downarrow_a\uparrow_b\rangle+|R_l^\downarrow\rangle|\downarrow_a\downarrow_b\rangle).
\end{eqnarray}
After $S$ gate is applied on the right arm, $|\Phi_{ep}\rangle_5$ is projected as
\begin{eqnarray}         \label{eq17}
\fl|\Phi_{ep}\rangle_6=
\frac{1}{2}\cos\alpha\cos\beta(|R_l^\downarrow\rangle|\uparrow_a\uparrow_b\rangle-|L_l^\uparrow\rangle|\uparrow_a\downarrow_b\rangle-|L_l^\uparrow\rangle|\downarrow_a\uparrow_b\rangle+|R_l^\downarrow\rangle|\downarrow_a\downarrow_b\rangle)\nonumber\\\fl\qquad\;\;\;-
\frac{1}{2}\cos\alpha\sin\beta(|L_r^\uparrow\rangle|\uparrow_a\uparrow_b\rangle+e^{i\pi/m}|R_r^\downarrow\rangle|\uparrow_a\downarrow_b\rangle-e^{i\pi/m}|R_r^\downarrow\rangle|\downarrow_a\uparrow_b\rangle-|L_r^\uparrow\rangle|\downarrow_a\downarrow_b\rangle)\nonumber\\\fl\qquad\;\;\;-
\frac{1}{2}\sin\alpha\cos\beta(|L_r^\uparrow\rangle|\uparrow_a\uparrow_b\rangle-e^{i\pi/m}|R_r^\downarrow\rangle|\uparrow_a\downarrow_b\rangle+e^{i\pi/m}|R_r^\downarrow\rangle|\downarrow_a\uparrow_b\rangle-|L_r^\uparrow\rangle|\downarrow_a\downarrow_b\rangle)\nonumber\\\fl\qquad\;\;\;+
\frac{1}{2}\sin\alpha\sin\beta(|R_l^\downarrow\rangle|\uparrow_a\uparrow_b\rangle+|L_l^\uparrow\rangle|\uparrow_a\downarrow_b\rangle+|L_l^\uparrow\rangle|\downarrow_a\uparrow_b\rangle+|R_l^\downarrow\rangle|\downarrow_a\downarrow_b\rangle).
\end{eqnarray}
BS induces $|\Phi_{ep}\rangle_6$ to be
\begin{eqnarray}         \label{eq18}
\fl|\Phi_{ep}\rangle_7=
\frac{1}{2\sqrt{2}}\cos\alpha\cos\beta(|R_l^\downarrow\rangle|\uparrow_a\uparrow_b\rangle-|L_l^\uparrow\rangle|\uparrow_a\downarrow_b\rangle-|L_l^\uparrow\rangle|\downarrow_a\uparrow_b\rangle+|R_l^\downarrow\rangle|\downarrow_a\downarrow_b\rangle)\nonumber\\\fl\qquad\;\;\;-
\frac{1}{2\sqrt{2}}\cos\alpha\sin\beta(|L_l^\uparrow\rangle|\uparrow_a\uparrow_b\rangle+e^{i\pi/m}|R_l^\downarrow\rangle|\uparrow_a\downarrow_b\rangle-e^{i\pi/m}|R_l^\downarrow\rangle|\downarrow_a\uparrow_b\rangle-|L_l^\uparrow\rangle|\downarrow_a\downarrow_b\rangle)\nonumber\\\fl\qquad\;\;\;-
\frac{1}{2\sqrt{2}}\sin\alpha\cos\beta(|L_l^\uparrow\rangle|\uparrow_a\uparrow_b\rangle-e^{i\pi/m}|R_l^\downarrow\rangle|\uparrow_a\downarrow_b\rangle+e^{i\pi/m}|R_l^\downarrow\rangle|\downarrow_a\uparrow_b\rangle-|L_l^\uparrow\rangle|\downarrow_a\downarrow_b\rangle)\nonumber\\\fl\qquad\;\;\;+
\frac{1}{2\sqrt{2}}\sin\alpha\sin\beta(|R_l^\downarrow\rangle|\uparrow_a\uparrow_b\rangle+|L_l^\uparrow\rangle|\uparrow_a\downarrow_b\rangle+|L_l^\uparrow\rangle|\downarrow_a\uparrow_b\rangle+|R_l^\downarrow\rangle|\downarrow_a\downarrow_b\rangle)\big)\nonumber\\\fl\qquad\;\;\;+
\frac{1}{2\sqrt{2}}\cos\alpha\cos\beta(|R_r^\downarrow\rangle|\uparrow_a\uparrow_b\rangle-|L_r^\uparrow\rangle|\uparrow_a\downarrow_b\rangle-|L_r^\uparrow\rangle|\downarrow_a\uparrow_b\rangle+|R_r^\downarrow\rangle|\downarrow_a\downarrow_b\rangle)\nonumber\\\fl\qquad\;\;\;+
\frac{1}{2\sqrt{2}}\cos\alpha\sin\beta(|L_r^\uparrow\rangle|\uparrow_a\uparrow_b\rangle+e^{i\pi/m}|R_r^\downarrow\rangle|\uparrow_a\downarrow_b\rangle-e^{i\pi/m}|R_r^\downarrow\rangle|\downarrow_a\uparrow_b\rangle-|L_r^\uparrow\rangle|\downarrow_a\downarrow_b\rangle)\nonumber\\\fl\qquad\;\;\;+
\frac{1}{2\sqrt{2}}\sin\alpha\cos\beta(|L_r^\uparrow\rangle|\uparrow_a\uparrow_b\rangle-e^{i\pi/m}|R_r^\downarrow\rangle|\uparrow_a\downarrow_b\rangle+e^{i\pi/m}|R_r^\downarrow\rangle|\downarrow_a\uparrow_b\rangle-|L_r^\uparrow\rangle|\downarrow_a\downarrow_b\rangle)\nonumber\\\fl\qquad\;\;\;+
\frac{1}{2\sqrt{2}}\sin\alpha\sin\beta(|R_r^\downarrow\rangle|\uparrow_a\uparrow_b\rangle+|L_r^\uparrow\rangle|\uparrow_a\downarrow_b\rangle+|L_r^\uparrow\rangle|\downarrow_a\uparrow_b\rangle+|R_r^\downarrow\rangle|\downarrow_a\downarrow_b\rangle).
\end{eqnarray}
$H_{ea}$ and $H_{eb}$ are performed on QD$_a$ and QD$_{b}$ again, which evolve $|\Phi_{ep}\rangle_7$ into
\begin{eqnarray}         \label{eq19}
\fl|\Phi_{ep}\rangle_8=
\frac{|+_l\rangle}{2}\big(
\cos\alpha\cos\beta|\downarrow_a\downarrow_b\rangle-
\cos\alpha\sin\beta(\frac{1-e^{i\pi/m}}{2}|\uparrow_a\downarrow_b\rangle+\frac{1+e^{i\pi/m}}{2}|\downarrow_a\uparrow_b\rangle)\nonumber\\\fl\qquad\;\;\;-
\sin\alpha\cos\beta(\frac{1+e^{i\pi/m}}{2}|\uparrow_a\downarrow_b\rangle+\frac{1-e^{i\pi/m}}{2}|\downarrow_a\uparrow_b\rangle)+
\sin\alpha\sin\beta|\uparrow_a\uparrow_b\rangle)\nonumber\\\fl\qquad\;\;\;+
\frac{|-_l\rangle}{2}\big(
\cos\alpha\cos\beta|\uparrow_a\uparrow_b\rangle+
\cos\alpha\sin\beta(\frac{1+e^{i\pi/m}}{2}|\uparrow_a\downarrow_b\rangle+\frac{1-e^{i\pi/m}}{2}|\downarrow_a\uparrow_b\rangle)\nonumber\\\fl\qquad\;\;\;+
\sin\alpha\cos\beta(\frac{1-e^{i\pi/m}}{2}|\uparrow_a\downarrow_b\rangle+\frac{1+e^{i\pi/m}}{2}|\downarrow_a\uparrow_b\rangle)+
\sin\alpha\sin\beta|\downarrow_a\downarrow_b\rangle)\nonumber\\\fl\qquad\;\;\;+
\frac{|+_r\rangle}{2}\big(
\cos\alpha\cos\beta|\downarrow_a\downarrow_b\rangle+
\cos\alpha\sin\beta(\frac{1-e^{i\pi/m}}{2}|\uparrow_a\downarrow_b\rangle+\frac{1+e^{i\pi/m}}{2}|\downarrow_a\uparrow_b\rangle)\nonumber\\\fl\qquad\;\;\;+
\sin\alpha\cos\beta(\frac{1+e^{i\pi/m}}{2}|\uparrow_a\downarrow_b\rangle+\frac{1-e^{i\pi/m}}{2}|\downarrow_a\uparrow_b\rangle)+
\sin\alpha\sin\beta|\uparrow_a\uparrow_b\rangle)\nonumber\\\fl\qquad\;\;\;+
\frac{|-_r\rangle}{2}\big(
\cos\alpha\cos\beta|\uparrow_a\uparrow_b\rangle
-\cos\alpha\sin\beta(\frac{1+e^{i\pi/m}}{2}|\uparrow_a\downarrow_b\rangle+\frac{1-e^{i\pi/m}}{2}|\downarrow_a\uparrow_b\rangle)\nonumber\\\fl\qquad\;\;\;
-\sin\alpha\cos\beta(\frac{1-e^{i\pi/m}}{2}|\uparrow_a\downarrow_b\rangle+\frac{1+e^{i\pi/m}}{2}|\downarrow_a\uparrow_b\rangle)+
\sin\alpha\sin\beta|\downarrow_a\downarrow_b\rangle).
\end{eqnarray}
Here $|\pm\rangle=(|R\rangle\pm|L\rangle)/\sqrt{2}$.

Lastly, we measure the outputting photon in the $\{|\pm\rangle\}$ basis, and apply some feed-forward operations on the QDs according to table \ref{table1}. And then, the joint state of the whole system is collapsed into
\begin{eqnarray}         \label{eq20}
\fl\qquad\;\;\;\;|\Phi_{e}\rangle_9=
\cos\alpha\cos\beta|\uparrow_a\uparrow_b\rangle+
\cos\alpha\sin\beta(\frac{1+e^{i\pi/m}}{2}|\uparrow_a\downarrow_b\rangle+\frac{1-e^{i\pi/m}}{2}|\downarrow_a\uparrow_b\rangle)\nonumber\\\fl\qquad\qquad\quad\;\;+
\sin\alpha\cos\beta(\frac{1-e^{i\pi/m}}{2}|\uparrow_a\downarrow_b\rangle+\frac{1+e^{i\pi/m}}{2}|\downarrow_a\uparrow_b\rangle)+
\sin\alpha\sin\beta|\downarrow_a\downarrow_b\rangle.
\end{eqnarray}
It is just the result of two--qubit $(\textrm{SWAP})^{1/m}$ gate on two electron--spin qubits $a$ and $b$. That is, the quantum circuit
shown in figure \ref{SWAP} can achieve a two-qubit $(\textrm{SWAP})^{1/m}$ gate in the two--qubit electron-spin system with a success probability of 100\% in principle.


\begin{table}[htb]
\centering \caption{The correlations between the outcomes of the outputting photon and the feed--forward operations for a  $\textrm{(SWAP)}^{1/m}$ gate with a success probability of 100\% in principle.  Here, $\sigma_{x}=|\uparrow\rangle\langle\downarrow|+|\downarrow\rangle\langle\uparrow|$, $\sigma_{z}=|\uparrow\rangle\langle\uparrow|-|\downarrow\rangle\langle\downarrow|$ and $I_2=|\uparrow\rangle\langle\uparrow|+|\downarrow\rangle\langle\downarrow|$.}
\begin{tabular}{ccc}
\hline  \hline

              & \multicolumn {2}{c}{Feed--forward} \\
\cline{2-3}
 $\;\;\;\;$    Detector     $\;\;\;\;$    &     $\;\;\;\;$      $\textrm{QD}_a$   $\;\;\;\;$    &   $\;\;\;\;$     $\textrm{QD}_b$   $\;\;\;\;$   \\
   \hline

$D_1$ $(|+_l\rangle)$  &     $\sigma_z\sigma_x$    &    $\sigma_z\sigma_x$     \\

$D_2$ $(|-_l\rangle)$  &     $I_2$                 &    $I_2$    \\

$D_3$ $(|+_r\rangle)$  &     $\sigma_x$            &    $\sigma_x$   \\

$D_4$ $(|-_r\rangle)$  &     $\sigma_z$            &    $\sigma_z$    \\

                             \hline  \hline
\end{tabular}\label{table1}
\end{table}

\section{Solid-state controlled--(swap)$^{1/m}$ gate} \label{sec3}

\begin{figure}[!h]
\begin{center}
\includegraphics[width=8.5 cm,angle=0]{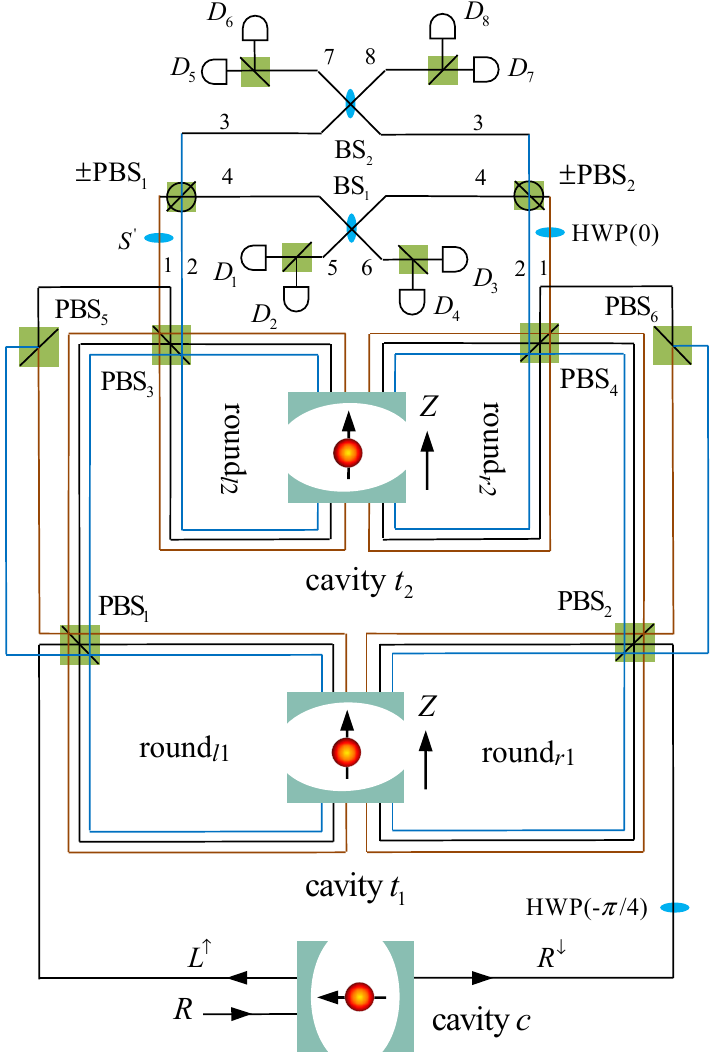}
\caption{(Color online) Schematic diagram for the construction of a deterministic controlled--(swap)$^{1/m}$ gate. HWP($-\pi/4$) represents HWP oriented at $-\pi/4$ degree to complete the transformation  $-|R\rangle\leftrightarrow|L\rangle$. HWP($0$) indicates HWP is rotated at 0 degree performing  the operation $\sigma_{z}=|R\rangle\langle R|-|L\rangle\langle L|$ on the passing photon. $S^{'}=|R\rangle\langle R|-e^{i\pi/m}|L\rangle\langle L|$ denotes phase shift operation acted on the passing photon.}
\label{Fredkin}
\end{center}
\end{figure}

The operation of the controlled--(swap)$^{1/m}$ gate can be written as
\begin{eqnarray}         \label{eq21}
U_{c-(\textrm{swap})^{1/m}}=\left(\begin{array}{cccccccc}
1  & 0 & 0  & 0 & 0 & 0 & 0 & 0 \\
0  & 1 & 0  & 0 & 0 & 0 & 0 & 0\\
0  & 0 & 1  & 0 & 0 & 0 & 0 & 0\\
0  & 0 & 0  & 1 & 0 & 0 & 0 & 0 \\
0  & 0 & 0  & 0 & 1 & 0 & 0 & 0 \\
0  & 0 & 0  & 0 & 0 &\frac{1+e^{i\pi/m}}{2} & \frac{1-e^{i\pi/m}}{2} & 0\\
0  & 0 & 0  & 0 & 0 & \frac{1-e^{i\pi/m}}{2} & \frac{1+e^{i\pi/m}}{2} & 0\\
0  & 0 & 0  & 0 & 0 & 0 & 0 & 1 \\
\end{array}\right).
\end{eqnarray}
The schematic diagram for implementing an electron-spin controlled-(swap)$^{1/m}$ gate is shown in figure \ref{Fredkin}, and it can be completed in the following five steps.

First, a single photon in the state $|R\rangle$ is injected. Based on the same arguments as made in above section, one can find that after the incident photon in the state $R$ interacts with QD$_c$, HWP($-\pi/4$), QD$_{t_1}$ and QD$_{t_2}$, the system evolutes from initial state
\begin{eqnarray}         \label{eq22}
|\varphi_{ep}\rangle_0&=
|R\rangle\otimes(\cos\alpha|\uparrow_c\rangle+\sin\alpha|\downarrow_c\rangle)\otimes
(\cos\beta|\uparrow_{t_1}\rangle+\sin\beta|\downarrow_{t_1}\rangle)\nonumber\\&\otimes
(\cos \delta|\uparrow_{t_2}\rangle+\sin\delta |\downarrow_{t_2}\rangle).
\end{eqnarray}
into
\begin{eqnarray}         \label{eq23}
|\varphi_{ep}\rangle_1=
\cos\alpha\cos\beta\cos\delta |L_r^\uparrow\rangle|\uparrow_c\uparrow_{t_1}\uparrow_{t_2}\rangle
-\cos\alpha\cos\beta\sin\delta |R_r^\downarrow\rangle|\uparrow_c\uparrow_{t_1}\downarrow_{t_2}\rangle\nonumber\\\qquad\;\;
-\cos\alpha\sin\beta\cos\delta |R_r^\downarrow\rangle|\uparrow_c\downarrow_{t_1}\uparrow_{t_2}\rangle
+\cos\alpha\sin\beta\sin\delta |L_r^\uparrow\rangle|\uparrow_c\downarrow_{t_1}\downarrow_{t_2}\rangle\nonumber\\\qquad\;\;
+\sin\alpha\cos\beta\cos\delta |L_l^\uparrow\rangle|\downarrow_c\uparrow_{t_1}\uparrow_{t_2}\rangle
-\sin\alpha\cos\beta\sin\delta |R_l^\downarrow\rangle|\downarrow_c\uparrow_{t_1}\downarrow_{t_2}\rangle\nonumber\\\qquad\;\;
-\sin\alpha\sin\beta\cos\delta |R_l^\downarrow\rangle|\downarrow_c\downarrow_{t_1}\uparrow_{t_2}\rangle
+\sin\alpha\sin\beta\sin\delta |L_l^\uparrow\rangle|\downarrow_c\downarrow_{t_1}\downarrow_{t_2}\rangle.
\end{eqnarray}
Here, HWP($-\pi/4$) represents a HWP set to $-\pi/4$ resulting in $-|R\rangle\leftrightarrow|L\rangle$.

Second, after $H_{et_1}$ and $H_{et_2}$  are performed on QD$_{t_1}$ and QD$_{t_2}$, PBS$_5$ leads $R_l^\downarrow$--polarized and $L_l^\uparrow$--polarized components to round$_{l1}$, round$_{l2}$, and $S'=|R\rangle\langle R|-e^{i\pi/m}|L\rangle\langle L|$ gate in succession. While PBS$_6$ guides $R_r^\downarrow$--polarized and  $L_r^\uparrow$-polarized components to round$_{r1}$, round$_{r2}$ and HWP(0) continuously.
Here, $S^{'}$ can be achieved by employing P$(\frac{1-m}{2m}\pi)$, QWP$(\frac{\pi}{4})$, HWP$(-\frac{\pi}{4m})$, QWP$(\frac{\pi}{4})$. HWP(0) realizes a $\sigma_{z}=|R\rangle\langle R|-|L\rangle\langle L|$ operation on the wave packets emitted from the $r_1$ arm.
Then, $|\varphi_{ep}\rangle_1$ is changed as
\begin{eqnarray}         \label{eq24}
\fl\qquad|\varphi_{ep}\rangle_2=\frac{1}{2}(
\cos\alpha\cos\beta\cos\delta|\uparrow_c\rangle(|L_{r_2}^\uparrow\rangle|\uparrow_{t_1}\uparrow_{t_2}\rangle-|R_{r_2}^\downarrow\rangle|\uparrow_{t_1}\downarrow_{t_2}\rangle-|R_{r_2}^\downarrow\rangle|\downarrow_{t_1}\uparrow_{t_2}\rangle\nonumber\\\fl\qquad\qquad\;\;+|L_{r_2}^\uparrow\rangle|\downarrow_{t_1}\downarrow_{t_2}\rangle)
\nonumber\\\fl\qquad\qquad\;\;
+\cos\alpha\cos\beta\sin\delta|\uparrow_c\rangle(-|R_{r_1}^\downarrow\rangle|\uparrow_{t_1}\uparrow_{t_2}\rangle+|L_{r_1}^\uparrow\rangle|\uparrow_{t_1}\downarrow_{t_2}\rangle-|L_{r_1}^\uparrow\rangle|\downarrow_{t_1}\uparrow_{t_2}\rangle\nonumber\\\fl\qquad\qquad\;\;+|R_{r_1}^\downarrow\rangle|\downarrow_{t_1}\downarrow_{t_2}\rangle)
\nonumber\\\fl\qquad\qquad\;\;
+\cos\alpha\sin\beta\cos\delta|\uparrow_c\rangle(-|R_{r_1}^\downarrow\rangle|\uparrow_{t_1}\uparrow_{t_2}\rangle-|L_{r_1}^\uparrow\rangle|\uparrow_{t_1}\downarrow_{t_2}\rangle+|L_{r_1}^\uparrow\rangle|\downarrow_{t_1}\uparrow_{t_2}\rangle\nonumber\\\fl\qquad\qquad\;\;+|R_{r_1}^\downarrow\rangle|\downarrow_{t_1}\downarrow_{t_2}\rangle)
\nonumber\\\fl\qquad\qquad\;\;
+\cos\alpha\sin\beta\sin\delta|\uparrow_c\rangle(|L_{r_2}^\uparrow\rangle|\uparrow_{t_1}\uparrow_{t_2}\rangle+|R_{r_2}^\downarrow\rangle|\uparrow_{t_1}\downarrow_{t_2}\rangle+|R_{r_2}^\downarrow\rangle|\downarrow_{t_1}\uparrow_{t_2}\rangle\nonumber\\\fl\qquad\qquad\;\;+|L_{r_2}^\uparrow\rangle|\downarrow_{t_1}\downarrow_{t_2}\rangle)
\nonumber\\\fl\qquad\qquad\;\;
+\sin\alpha\cos\beta\cos\delta|\downarrow_c\rangle(|L_{l_2}^\uparrow\rangle|\uparrow_{t_1}\uparrow_{t_2}\rangle-|R_{l_2}^\downarrow\rangle|\uparrow_{t_1}\downarrow_{t_2}\rangle-|R_{l_2}^\downarrow\rangle|\downarrow_{t_1}\uparrow_{t_2}\rangle\nonumber\\\fl\qquad\qquad\;\;+|L_{l_2}^\uparrow\rangle|\downarrow_{t_1}\downarrow_{t_2}\rangle)
\nonumber\\\fl\qquad\qquad\;\;
+\sin\alpha\cos\beta\sin\delta|\downarrow_c\rangle(-|R_{l_1}^\downarrow\rangle|\uparrow_{t_1}\uparrow_{t_2}\rangle+e^{i\pi/m}|L_{l_1}^\uparrow\rangle|\uparrow_{t_1}\downarrow_{t_2}\rangle-e^{i\pi/m}|L_{l_1}^\uparrow\rangle|\downarrow_{t_1}\uparrow_{t_2}\rangle\nonumber\\\fl\qquad\qquad\;\;+|R_{l_1}^\downarrow\rangle|\downarrow_{t_1}\downarrow_{t_2}\rangle)
\nonumber\\\fl\qquad\qquad\;\;
+\sin\alpha\sin\beta\cos\delta|\downarrow_c\rangle(-|R_{l_1}^\downarrow\rangle|\uparrow_{t_1}\uparrow_{t_2}\rangle-e^{i\pi/m}|L_{l_1}^\uparrow\rangle|\uparrow_{t_1}\downarrow_{t_2}\rangle+e^{i\pi/m}|L_{l_1}^\uparrow\rangle|\downarrow_{t_1}\uparrow_{t_2}\rangle\nonumber\\\fl\qquad\qquad\;\;+|R_{l_1}^\downarrow\rangle|\downarrow_{t_1}\downarrow_{t_2}\rangle)
\nonumber\\\fl\qquad\qquad\;\;
+\sin\alpha\sin\beta\sin\delta|\downarrow_c\rangle(|L_{l_2}^\uparrow\rangle|\uparrow_{t_1}\uparrow_{t_2}\rangle+|R_{l_2}^\downarrow\rangle|\uparrow_{t_1}\downarrow_{t_2}\rangle+|R_{l_2}^\downarrow\rangle|\downarrow_{t_1}\uparrow_{t_2}\rangle\nonumber\\\fl\qquad\qquad\;\;+|L_{l_2}^\uparrow\rangle|\downarrow_{t_1}\downarrow_{t_2}\rangle) ).
\end{eqnarray}

Third, after the wave packets pass through $\pm$PBS$_1$ and $\pm$PBS$_2$, Hadamard operations $H_{et_1}$ and $H_{et_2}$  are applied on QD$_{t_1}$ and QD$_{t_2}$, respectively.  Here $\pm$PBS$_1$, $\pm$PBS$_2$, $H_{et_1}$ and $H_{et_2}$ transform state of the system into
\begin{eqnarray}         \label{eq25}
\fl\qquad|\varphi_{ep}\rangle_3= \frac{1}{\sqrt{2}} (
 |+_{r_3}\rangle\cos\alpha\cos\beta\cos\delta|\uparrow_c\downarrow_{t_1}\downarrow_{t_2}\rangle
-|-_{r_3}\rangle\cos\alpha\cos\beta\sin\delta|\uparrow_c\downarrow_{t_1}\uparrow_{t_2}\rangle\nonumber\\\fl\qquad\qquad\;\;
-|-_{r_3}\rangle\cos\alpha\sin\beta\cos\delta|\uparrow_c\uparrow_{t_1}\downarrow_{t_2}\rangle
+|+_{r_3}\rangle\cos\alpha\sin\beta\sin\delta|\uparrow_c\uparrow_{t_1}\uparrow_{t_2}\rangle\nonumber\\\fl\qquad\qquad\;\;
-|-_{r_4}\rangle\cos\alpha\cos\beta\cos\delta|\uparrow_c\uparrow_{t_1}\uparrow_{t_2}\rangle
-|+_{r_4}\rangle\cos\alpha\cos\beta\sin\delta|\uparrow_c\uparrow_{t_1}\downarrow_{t_2}\rangle\nonumber\\\fl\qquad\qquad\;\;
-|+_{r_4}\rangle\cos\alpha\sin\beta\cos\delta|\uparrow_c\downarrow_{t_1}\uparrow_{t_2}\rangle
-|-_{r_4}\rangle\cos\alpha\sin\beta\sin\delta|\uparrow_c\downarrow_{t_1}\downarrow_{t_2}\rangle\nonumber\\\fl\qquad\qquad\;\;
+|+_{l_3}\rangle\sin\alpha\cos\beta\cos\delta|\downarrow_c\downarrow_{t_1}\downarrow_{t_2}\rangle\nonumber\\\fl\qquad\qquad\;\;
-|-_{l_3}\rangle\sin\alpha\cos\beta\sin\delta(\frac{1-e^{i\pi/m}}{2}|\downarrow_c\uparrow_{t_1}\downarrow_{t_2}\rangle+\frac{1+e^{i\pi/m}}{2}|\downarrow_c\downarrow_{t_1}\uparrow_{t_2}\rangle)\nonumber\\\fl\qquad\qquad\;\;
-|-_{l_3}\rangle\sin\alpha\sin\beta\cos\delta(\frac{1+e^{i\pi/m}}{2}|\downarrow_c\uparrow_{t_1}\downarrow_{t_2}\rangle+\frac{1-e^{i\pi/m}}{2}|\downarrow_c\downarrow_{t_1}\uparrow_{t_2}\rangle)\nonumber\\\fl\qquad\qquad\;\;
+|+_{l_3}\rangle\sin\alpha\sin\beta\sin\delta|\downarrow_c\uparrow_{t_1}\uparrow_{t_2}\rangle
-|-_{l_4}\rangle\sin\alpha\cos\beta\cos\delta|\downarrow_c\uparrow_{t_1}\uparrow_{t_2}\rangle\nonumber\\\fl\qquad\qquad\;\;
-|+_{l_4}\rangle\sin\alpha\cos\beta\sin\delta(\frac{1+e^{i\pi/m}}{2}|\downarrow_c\uparrow_{t_1}\downarrow_{t_2}\rangle+\frac{1-e^{i\pi/m}}{2}|\downarrow_c\downarrow_{t_1}\uparrow_{t_2}\rangle)\nonumber\\\fl\qquad\qquad\;\;
-|+_{l_4}\rangle\sin\alpha\sin\beta\cos\delta(\frac{1-e^{i\pi/m}}{2}|\downarrow_c\uparrow_{t_1}\downarrow_{t_2}\rangle+\frac{1+e^{i\pi/m}}{2}|\downarrow_c\downarrow_{t_1}\uparrow_{t_2}\rangle)\nonumber\\\fl\qquad\qquad\;\;
-|-_{l_4}\rangle\sin\alpha\sin\beta\sin\delta|\downarrow_c\downarrow_{t_1}\downarrow_{t_2}\rangle ).
\end{eqnarray}

Fourth, after the wave packets emitted from the spatial $l_4$ ($l_3$) converge with the wave packets emitted from $r_4$ ($r_3$) at 50:50 BS$_1$ (BS$_2$), $|\varphi_{ep}\rangle_3$ is converted into
\begin{eqnarray}         \label{eq26}
\fl\qquad|\varphi_{ep}\rangle_4= \frac{1}{2} (
 |+_{7}\rangle\cos\alpha\cos\beta\cos\delta|\uparrow_c\downarrow_{t_1}\downarrow_{t_2}\rangle
-|-_{7}\rangle\cos\alpha\cos\beta\sin\delta|\uparrow_c\downarrow_{t_1}\uparrow_{t_2}\rangle\nonumber\\\fl\qquad\qquad\;\;
-|-_{7}\rangle\cos\alpha\sin\beta\cos\delta|\uparrow_c\uparrow_{t_1}\downarrow_{t_2}\rangle
+|+_{7}\rangle\cos\alpha\sin\beta\sin\delta|\uparrow_c\uparrow_{t_1}\uparrow_{t_2}\rangle\nonumber\\\fl\qquad\qquad\;\;
+|+_{7}\rangle\sin\alpha\cos\beta\cos\delta|\downarrow_c\downarrow_{t_1}\downarrow_{t_2}\rangle\nonumber\\\fl\qquad\qquad\;\;
-|-_{7}\rangle\sin\alpha\cos\beta\sin\delta(\frac{1-e^{i\pi/m}}{2}|\downarrow_c\uparrow_{t_1}\downarrow_{t_2}\rangle+\frac{1+e^{i\pi/m}}{2}|\downarrow_c\downarrow_{t_1}\uparrow_{t_2}\rangle)\nonumber\\\fl\qquad\qquad\;\;
-|-_{7}\rangle\sin\alpha\sin\beta\cos\delta(\frac{1+e^{i\pi/m}}{2}|\downarrow_c\uparrow_{t_1}\downarrow_{t_2}\rangle+\frac{1-e^{i\pi/m}}{2}|\downarrow_c\downarrow_{t_1}\uparrow_{t_2}\rangle)\nonumber\\\fl\qquad\qquad\;\;
+|+_{7}\rangle\sin\alpha\sin\beta\sin\delta|\downarrow_c\uparrow_{t_1}\uparrow_{t_2}\rangle
-|+_{8}\rangle\cos\alpha\cos\beta\cos\delta|\uparrow_c\downarrow_{t_1}\downarrow_{t_2}\rangle\nonumber\\\fl\qquad\qquad\;\;
+|-_{8}\rangle\cos\alpha\cos\beta\sin\delta|\uparrow_c\downarrow_{t_1}\uparrow_{t_2}\rangle
+|-_{8}\rangle\cos\alpha\sin\beta\cos\delta|\uparrow_c\uparrow_{t_1}\downarrow_{t_2}\rangle\nonumber\\\fl\qquad\qquad\;\;
-|+_{8}\rangle\cos\alpha\sin\beta\sin\delta|\uparrow_c\uparrow_{t_1}\uparrow_{t_2}\rangle
+|+_{8}\rangle\sin\alpha\cos\beta\cos\delta|\downarrow_c\downarrow_{t_1}\downarrow_{t_2}\rangle\nonumber\\\fl\qquad\qquad\;\;
-|-_{8}\rangle\sin\alpha\cos\beta\sin\delta(\frac{1-e^{i\pi/m}}{2}|\downarrow_c\uparrow_{t_1}\downarrow_{t_2}\rangle+\frac{1+e^{i\pi/m}}{2}|\downarrow_c\downarrow_{t_1}\uparrow_{t_2}\rangle)\nonumber\\\fl\qquad\qquad\;\;
-|-_{8}\rangle\sin\alpha\sin\beta\cos\delta(\frac{1+e^{i\pi/m}}{2}|\downarrow_c\uparrow_{t_1}\downarrow_{t_2}\rangle+\frac{1-e^{i\pi/m}}{2}|\downarrow_c\downarrow_{t_1}\uparrow_{t_2}\rangle)\nonumber\\\fl\qquad\qquad\;\;
+|+_{8}\rangle\sin\alpha\sin\beta\sin\delta|\downarrow_c\uparrow_{t_1}\uparrow_{t_2}\rangle
-|-_{5}\rangle\cos\alpha\cos\beta\cos\delta|\uparrow_c\uparrow_{t_1}\uparrow_{t_2}\rangle\nonumber\\\fl\qquad\qquad\;\;
-|+_{5}\rangle\cos\alpha\cos\beta\sin\delta|\uparrow_c\uparrow_{t_1}\downarrow_{t_2}\rangle
-|+_{5}\rangle\cos\alpha\sin\beta\cos\delta|\uparrow_c\downarrow_{t_1}\uparrow_{t_2}\rangle\nonumber\\\fl\qquad\qquad\;\;
-|-_{5}\rangle\cos\alpha\sin\beta\sin\delta|\uparrow_c\downarrow_{t_1}\downarrow_{t_2}\rangle
-|-_{5}\rangle\sin\alpha\cos\beta\cos\delta|\downarrow_c\uparrow_{t_1}\uparrow_{t_2}\rangle\nonumber\\\fl\qquad\qquad\;\;
-|+_{5}\rangle\sin\alpha\cos\beta\sin\delta(\frac{1+e^{i\pi/m}}{2}|\downarrow_c\uparrow_{t_1}\downarrow_{t_2}\rangle+\frac{1-e^{i\pi/m}}{2}|\downarrow_c\downarrow_{t_1}\uparrow_{t_2}\rangle)\nonumber\\\fl\qquad\qquad\;\;
-|+_{5}\rangle\sin\alpha\sin\beta\cos\delta(\frac{1-e^{i\pi/m}}{2}|\downarrow_c\uparrow_{t_1}\downarrow_{t_2}\rangle+\frac{1+e^{i\pi/m}}{2}|\downarrow_c\downarrow_{t_1}\uparrow_{t_2}\rangle)\nonumber\\\fl\qquad\qquad\;\;
-|-_{5}\rangle\sin\alpha\sin\beta\sin\delta|\downarrow_c\downarrow_{t_1}\downarrow_{t_2}\rangle
+|-_{6}\rangle\cos\alpha\cos\beta\cos\delta|\uparrow_c\uparrow_{t_1}\uparrow_{t_2}\rangle\nonumber\\\fl\qquad\qquad\;\;
+|+_{6}\rangle\cos\alpha\cos\beta\sin\delta|\uparrow_c\uparrow_{t_1}\downarrow_{t_2}\rangle
+|+_{6}\rangle\cos\alpha\sin\beta\cos\delta|\uparrow_c\downarrow_{t_1}\uparrow_{t_2}\rangle\nonumber\\\fl\qquad\qquad\;\;
+|-_{6}\rangle\cos\alpha\sin\beta\sin\delta|\uparrow_c\downarrow_{t_1}\downarrow_{t_2}\rangle
-|-_{6}\rangle\sin\alpha\cos\beta\cos\delta|\downarrow_c\uparrow_{t_1}\uparrow_{t_2}\rangle\nonumber\\\fl\qquad\qquad\;\;
-|+_{6}\rangle\sin\alpha\cos\beta\sin\delta(\frac{1+e^{i\pi/m}}{2}|\downarrow_c\uparrow_{t_1}\downarrow_{t_2}\rangle+\frac{1-e^{i\pi/m}}{2}|\downarrow_c\downarrow_{t_1}\uparrow_{t_2}\rangle)\nonumber\\\fl\qquad\qquad\;\;
-|+_{6}\rangle\sin\alpha\sin\beta\cos\delta(\frac{1-e^{i\pi/m}}{2}|\downarrow_c\uparrow_{t_1}\downarrow_{t_2}\rangle+\frac{1+e^{i\pi/m}}{2}|\downarrow_c\downarrow_{t_1}\uparrow_{t_2}\rangle)\nonumber\\\fl\qquad\qquad\;\;
-|-_{6}\rangle\sin\alpha\sin\beta\sin\delta|\downarrow_c\downarrow_{t_1}\downarrow_{t_2}\rangle ).
\end{eqnarray}

Fifth, from equation (\ref{eq26}), one can see that after detecting the output photon in the $\{|R\rangle,\; |L\rangle\}$ basis and applying some proper feed--forward operations on QD$_c$, QD$_{t_1}$, and QD$_{t_2}$ (see table \ref{table2}), the state of the whole system will collapse into
\begin{eqnarray}         \label{eq28}
\fl\qquad\qquad\;\;|\varphi_{e}\rangle_5=
\cos\alpha\cos\beta\cos\delta|\uparrow_c\uparrow_{t_1}\uparrow_{t_2}\rangle
+\cos\alpha\cos\beta\sin\delta|\uparrow_c\uparrow_{t_1}\downarrow_{t_2}\rangle\nonumber\\\fl\qquad\qquad\qquad\;\;\;
+\cos\alpha\sin\beta\cos\delta|\uparrow_c\downarrow_{t_1}\uparrow_{t_2}\rangle
+\cos\alpha\sin\beta\sin\delta|\uparrow_c\downarrow_{t_1}\downarrow_{t_2}\rangle\nonumber\\\fl\qquad\qquad\qquad\;\;\;
+\sin\alpha\cos\beta\cos\delta|\downarrow_c\uparrow_{t_1}\uparrow_{t_2}\rangle\nonumber\\\fl\qquad\qquad\qquad\;\;\;
+\sin\alpha\cos\beta\sin\delta(\frac{1+e^{i\pi/m}}{2}|\downarrow_c\uparrow_{t_1}\downarrow_{t_2}\rangle+\frac{1-e^{i\pi/m}}{2}|\downarrow_c\downarrow_{t_1}\uparrow_{t_2}\rangle)\nonumber\\\fl\qquad\qquad\qquad\;\;\;
+\sin\alpha\sin\beta\cos\delta(\frac{1-e^{i\pi/m}}{2}|\downarrow_c\uparrow_{t_1}\downarrow_{t_2}\rangle+\frac{1+e^{i\pi/m}}{2}|\downarrow_c\downarrow_{t_1}\uparrow_{t_2}\rangle)\nonumber\\\fl\qquad\qquad\qquad\;\;\;
+\sin\alpha\sin\beta\sin\delta|\downarrow_c\downarrow_{t_1}\downarrow_{t_2}\rangle.
\end{eqnarray}

Therefore, the quantum circuit shown in figure \ref{Fredkin} can accomplish a deterministic controlled--(swap)$^{1/m}$ gate which implements a (SWAP)$^{1/m}$ operation on two target QD--spin qubits if and only if the control QD--spin  qubit is in the state $|\downarrow\rangle$.

\begin{table}[htb]
\centering
\caption{The relations between the outcomes of the measurements on the photon and the feed--forward operations for a deterministic controlled--(swap)$^{1/m}$ gate.}
\begin{tabular}{cccccc}
\hline  \hline

          & \multicolumn {3}{c}{Feed--forward} \\
\hline
\;\; Detector       &   QD$_c$      &   QD$_{t_1}$    &    QD$_{t_2}$ \\

$D_1$    &  $I_2$         &  $I_2$         &   $I_2$   \\

$D_2$    &  $I_2$         & $\sigma_z$     &  $\sigma_z$      \\

$D_3$    &  $\sigma_z$     & $I_2$         &   $ I_2$  \\

$D_4$    &   $\sigma_z$    &  $\sigma_z$    &   $\sigma_z$    \\

$D_5$    &  $I_2$           &  $\sigma_z$$\sigma_x$    &   $\sigma_z$$\sigma_x$  \\

$D_6$    &  $I_2$           &  $\sigma_x$          &   $\sigma_x$
     \\

$D_7$    &  $\sigma_z$      & $\sigma_z$$\sigma_x$   &   $ \sigma_z$$\sigma_x$   \\

$D_8$   &  $\sigma_z$        &  $\sigma_x$           &   $\sigma_x$     \\
                             \hline  \hline
\end{tabular}\label{table2}
\end{table}

\section{Discussion and conclusion} \label{sec3}

In an ideal case, the photon--QD emitter will work deterministically and without any experimental errors. However, the experimental performance of the photon--cavity emitter (equation (\ref{eq2})) may decrease by a factor of $1-\exp(-\tau/T_2)$ due to exciton dephasing. Here, $\tau=1/g$ is the cavity photon lifetime (Rabi oscillation period). $T_2\sim \mu$s is the electron spin coherence time, which is limited by the spin-relaxation time $T_1\sim$ ms. The optical dephasing has a slight impact on the photon--QD emitter because the optical coherence time \cite{optical-time} (several hundred picoseconds \cite{picoseconds2}) is ten times longer than the cavity photon lifetime $\tau$ (tens of picoseconds \cite{picoseconds3}).
The hole spin coherence time ($T_2^h>100$ ns) is three orders of magnitude longer than the cavity photon lifetime\cite{picoseconds2,picoseconds3,picoseconds4}, causing the spin dephasing to be safely neglected.
Moreover, the fidelity of the emitter might be reduced by a few percent as a result of the heavy--light hole mixing generated by imperfect optical selection rules \cite{impefer1}. Fortunately, the heavy--light hole mixing could be decreased by improving the shape, size, and type of charged exciton\cite{decrease1,decrease2}.
In  experiment, the coupling loss and mismatch between the photon propagating mode and the cavity mode will decrease the fidelity of the emitter. The quantum emitter loss mechanisms in the few-photon regime have been investigated, and great progress has been made in scenarios of the scattering process \cite{emitter-loss1}, material absorption \cite{emitter-loss2}, and spontaneous emissions \cite{emitter-loss3}.

The spin-dependent optical rules shown in equation (\ref{eq4}) are the key ingredients in implementing our (SWAP)$^{1/m}$ and controlled--(swap)$^{1/m}$ gates. In practice, the imperfect birefringence and side leakage from the cavity are unavoidable in the experiment\cite{leakage1,leakage2}, and they decrease the performance of the photon--QD emitter. We took into account  two such imperfections with the resonant condition $\omega_{X^-}=\omega_c=\omega$;  $t_0$ and $r_0$ are described by equation (\ref{eq3}) with $\textrm{g}=0$. $t_0$ and $r$, $t$ and $r_0$ have opposite signs. Thus, equation (\ref{eq4}) can be modified as
\begin{eqnarray}       \label{eq27}
&|R^\uparrow   \uparrow\rangle \rightarrow  r|L^\downarrow \uparrow\rangle  +t|R^\uparrow \uparrow\rangle,\quad
|R^\uparrow   \downarrow\rangle \rightarrow t_0|R^\uparrow \downarrow\rangle + r_0|L^\downarrow \downarrow\rangle,\nonumber\\\;
&|L^\downarrow \uparrow\rangle \rightarrow  r|R^\uparrow \uparrow\rangle  +t|L^\downarrow \uparrow\rangle,\;\;\;\;
|L^\downarrow\downarrow\rangle \rightarrow t_0|L^\downarrow\downarrow\rangle + r_0|R^\uparrow\downarrow\rangle,\nonumber\\
&|L^\uparrow  \downarrow\rangle \rightarrow r|R^\downarrow  \downarrow\rangle +t|L^\uparrow,\downarrow\rangle,\quad
|R^\downarrow \uparrow\rangle \rightarrow t_0|R^\downarrow \uparrow\rangle  +r_0|L^\uparrow \uparrow\rangle,\nonumber\\
&|R^\downarrow \downarrow\rangle \rightarrow r|L^\uparrow \downarrow\rangle +t|R^\downarrow \downarrow\rangle,\quad
|L^\uparrow   \uparrow\rangle \rightarrow t_0|L^\uparrow \uparrow\rangle  +r_0|R^\downarrow \uparrow\rangle.
\end{eqnarray}


To evaluate the performance of the (SWAP)$^{1/m}$  and controlled-(swap)$^{1/m}$ gates, we calculated the average fidelities and efficiencies of the two gates.  In the experiment, the fidelity and efficiency of the gates were defined by $F=|\langle \psi_f|\psi_i\rangle|^2$ and $\eta=\frac{n_{\rm{output}}}{n_{\rm{input}}}$, respectively. Here, $|\psi_i\rangle$ and $|\psi_f\rangle$ represent the ideal (equation (\ref{eq4})) and practical output states (equation (\ref{eq27})), respectively. $n_{\rm{input}}$ and $n_{\rm{output}}$ are the number of detecting photons before and after the schemes, respectively.  Hence,  the average fidelities and efficiencies can be written as
\begin{eqnarray}       \label{eq28}
\overline{F}_{{\rm(SWAP)}^{1/m}}=&\frac{1}{(2\pi)^2}\int_{0}^{2\pi}d\alpha\int_{0}^{2\pi}d\beta |\langle \Phi_f|\Phi_i\rangle|^2,
\end{eqnarray}
\begin{eqnarray}       \label{eq29}
\overline{F}_{{\rm c-(swap)}^{1/m}}=&\frac{1}{(2\pi)^3}\int_{0}^{2\pi}d\alpha\int_{0}^{2\pi}d\beta\int_{0}^{2\pi}d\delta \;   |\langle \varphi_f|\varphi_i\rangle|^2,
\end{eqnarray}
\begin{eqnarray}       \label{eq30}
\overline{\eta}_{{\rm(SWAP)}^{1/m}}=&\frac{1}{(2\pi)^2}\int_{0}^{2\pi}d\alpha\int_{0}^{2\pi}d\beta \frac{n_{\rm{output}}}{n_{\rm{intput}}},
\end{eqnarray}
\begin{eqnarray}       \label{eq31}
\overline{\eta}_{{\rm c-(swap)}^{1/m}}=&\frac{1}{(2\pi)^3}\int_{0}^{2\pi}d\alpha\int_{0}^{2\pi}d\beta\int_{0}^{2\pi}d\delta \; \frac{n_{\rm{output}}}{n_{\rm{intput}}}.
\end{eqnarray}

\begin{figure}[!h]
\centering
\includegraphics[width=7.50 cm,angle=0]{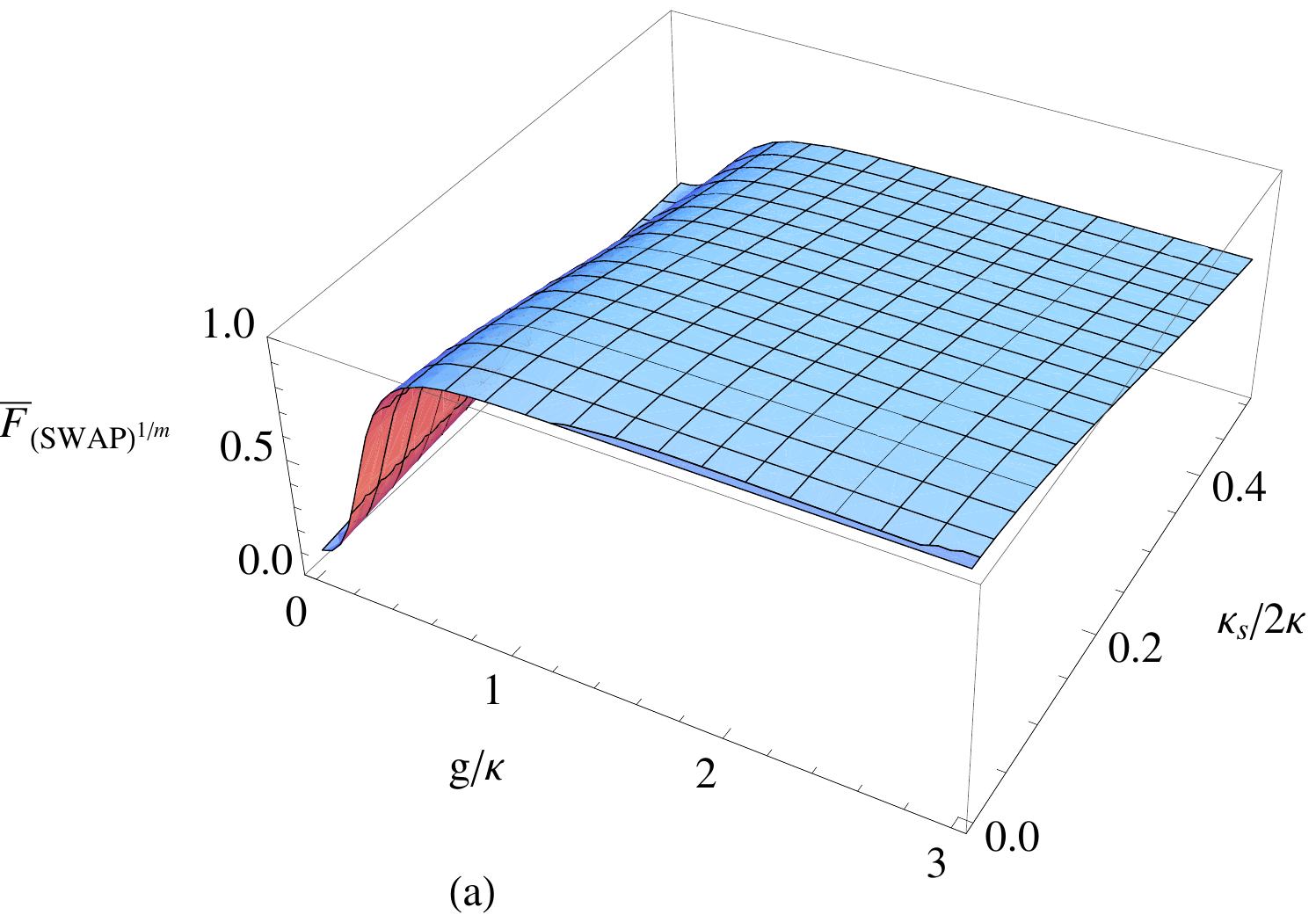}
\includegraphics[width=7.50 cm,angle=0]{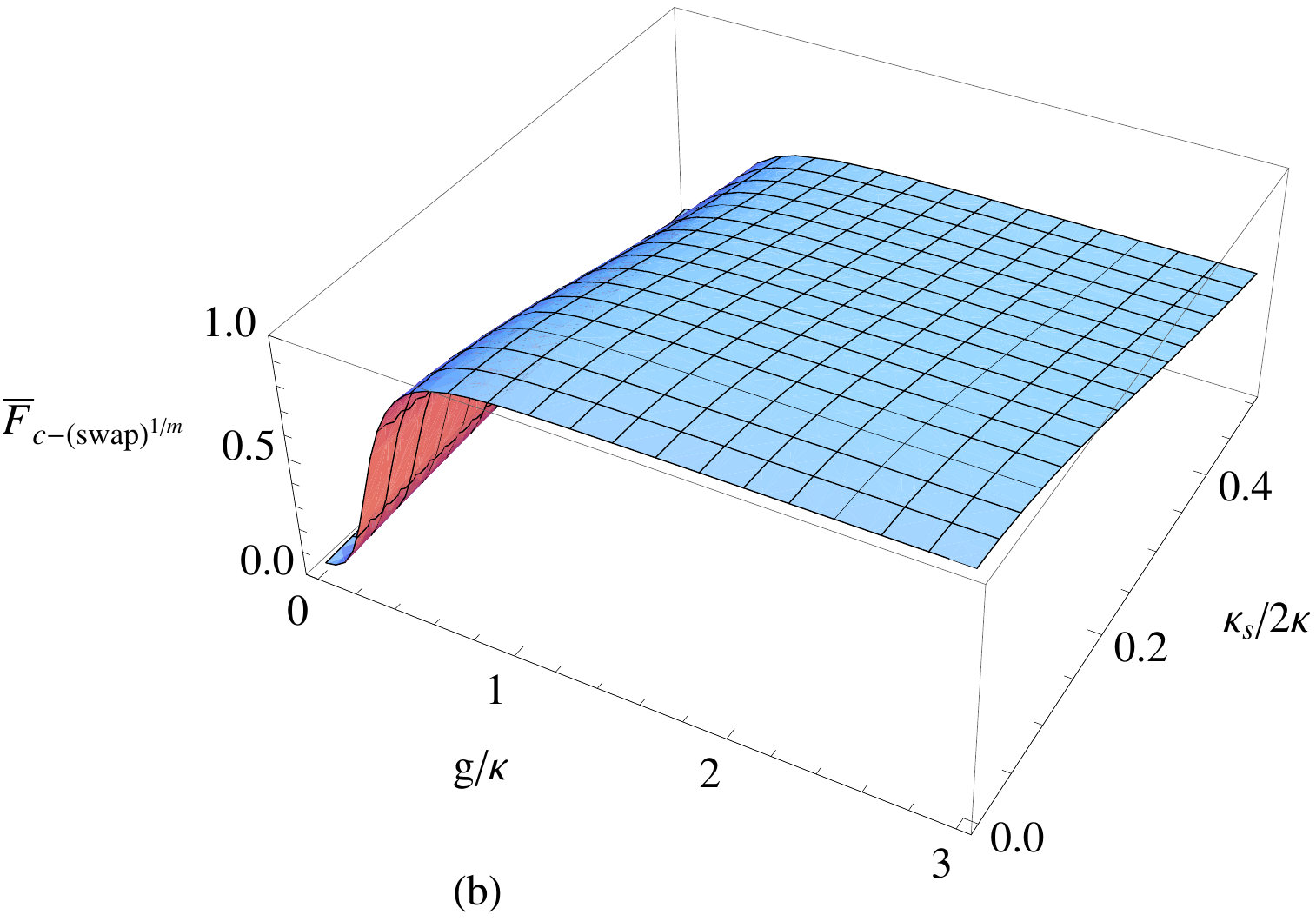}
\caption{(Color online)  The average fidelities of the two more general solid--state gates as functions of $g/\kappa$ and $\kappa_s/2\kappa$. (a) The average fidelity of the (SWAP)$^{1/m}$ gate; (b) The average fidelity of the controlled-(swap)$^{1/m}$ gate. $w_c=\omega_{X^-}=\omega$, and $\gamma=0.1\kappa$ are taken.} \label{Fidelity}
\end{figure}

\begin{figure}  [!h]        
\centering
\includegraphics[width=7.50 cm,angle=0]{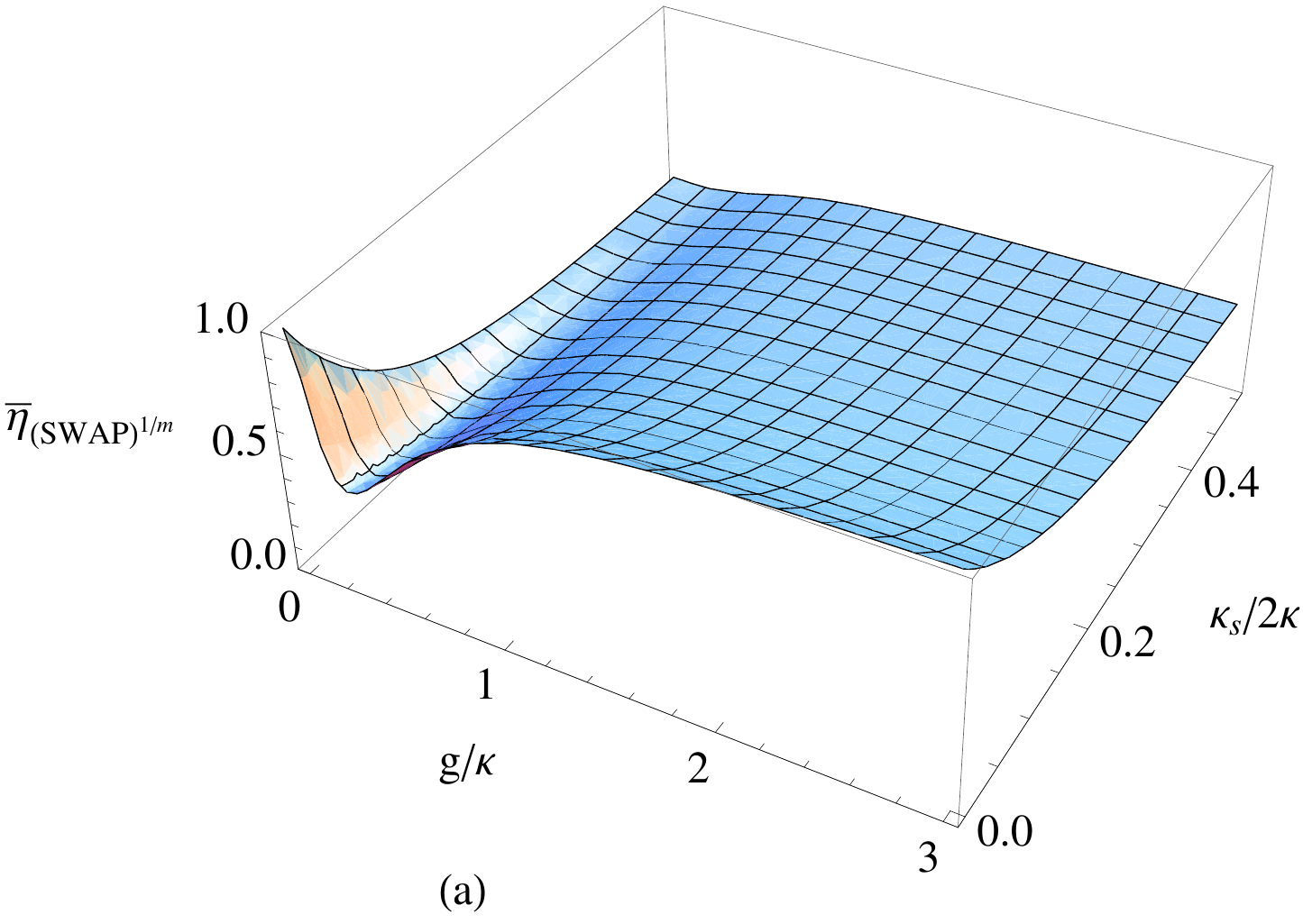}
\includegraphics[width=7.50 cm,angle=0]{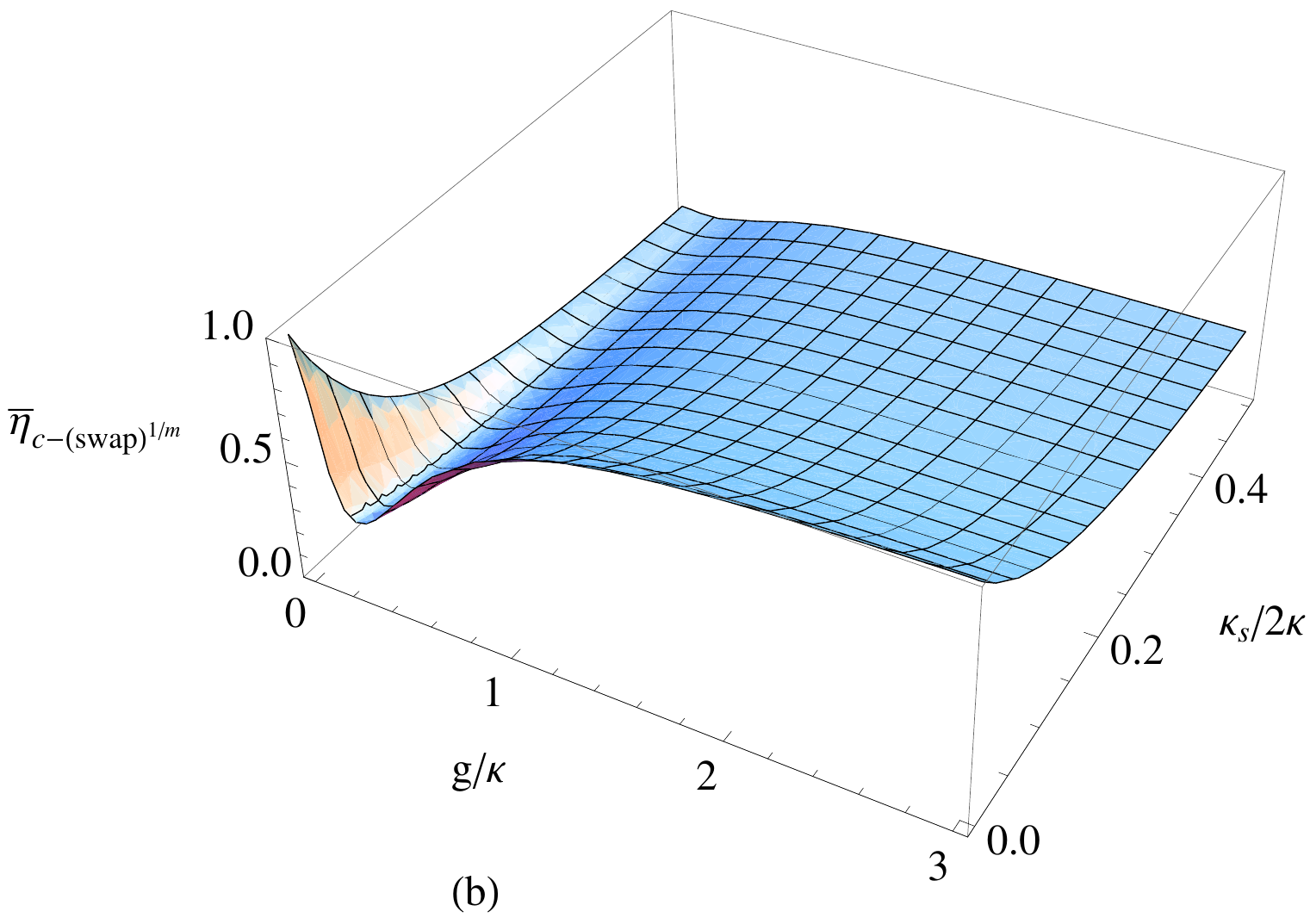}\\
\caption{(Color online)   The average efficiencies of the proposed two more general solid-state gates as functions of $g/\kappa$ and $\kappa_s/2\kappa$. (a) The average efficiency of the (SWAP)$^{1/m}$ gate; (b) The average efficiency of the controlled--(swap)$^{1/m}$ gate. $w_c=\omega_{X^-}=\omega$, and $\gamma=0.1\kappa$ are taken. } \label{Efficiency}
\end{figure}

Without considering the coupling loss between the photon-propagating mode and the cavity mode, the average fidelities and efficiencies of the two gates are functions of $g/\kappa$ and $\kappa_s/2\kappa$, respectively,  and are described in figure \ref{Fidelity} and figure \ref{Efficiency}, respectively. It is observed that the cavity side leakage, imperfect birefringence, and QD--cavity coupling strength had a great impact on the fidelities and efficiencies.
When $\kappa_s\ll\kappa$, our schemes work efficiently, even in weak coupling regimes.
When $\kappa_s>\kappa$ is satisfied, efficient joining schemes work only in strong coupling regimes defined by $g>(\kappa+\kappa_s)/4$.
For example, $g/(\kappa+\kappa_s)=2.7$  and $\kappa_s/\kappa=0.05$  \cite{Strong-coupling1} are available, and in this case, the average fidelities of (SWAP)$^{1/m}$ and controlled--(swap)$^{1/m}$ gates are  99.09\% and  98.93\%, respectively. If $g/(\kappa+\kappa_s)=0.5$ and $\kappa_s/\kappa=0$, the average fidelities are approximately 89.36\%  and  87.42\%, respectively. When $g/(\kappa+\kappa_s)=2.5$ and $\kappa_s/\kappa=0$, the average fidelities are approximately 99.98\% and 99.97\%, respectively, and the average efficiency are both over 96.13\%.

Strong coupling is a challenge in practice.  Fortunately, in 2004,  $\kappa_s/\kappa$ was reduced to $0.7$ with $g/(\kappa+\kappa_s)=1$ \cite{Strong-coupling1,Strong-coupling2,Strong-coupling3}.  The coupling strength $g/(\kappa+\kappa_s)$ was raised from 0.5 ($Q = 8800$) \cite{Strong-coupling1} to 2.4 ($Q = 40000$) \cite{Q=40000} in 1.5 $\mu m$ micropillar microcavities by improving the sample design, growth, and fabrication \cite{Strong-coupling2}. $g/(\kappa+\kappa_s)=2.7$ was reported for a single InGaAs QD in 2012 \cite{g=2.7}.
$\kappa_s/\kappa\approx0.2$ was reported for QD-pillar cavity in 2012 \cite{ks/k=0.2}.
It has been reported that $\kappa_s/\kappa=0.05$ in the strong regime could be achieved in a pillar microcavity with $Q = 9000$ \cite{Strong-coupling1}. As quality factor $Q$ increases, the side leakage rate $\kappa_s$ may decrease \cite{ks-small}. Experiments have shown that the existence of strong coupling is registered to a single QD with a coupling rate attaining to 120 meV \cite{strongcoupling0,strongcoupling00}. Strong coupling of QD emitters has had major breakthroughs \cite{strongcouplingQD}. Furthermore, coupling strength variations have been explained in diverse structures of a single QD at room temperature, which provides a possible solution for the realization of strong coupling \cite{strongcoupling}.

To summarize, we have proposed two deterministic schemes to implement solid--state more general (SWAP)$^{1/m}$ and controlled--(swap)$^{1/m}$  gates for integer $m \geq1$. Particularly, one can obtain SWAP and Fredkin gates  for $m=1$; one can also acquire (SWAP)$^{1/2}$ for $m=2$. The long--distance gate qubits are bridged by  flying single--photon mediate. The flexible parameter $m$ in these gates can be adjusted by using two QWPs and one HWP.
It is known that at least six photon--matter interactions are required to implement a photonic or matter SWAP gate, that the optimal synthesis of a SWAP gate is three CNOT gates \cite{book}, and that one or two photon--matter interactions are not sufficient for implementing a CNOT gate.
Two parity--check gates and one additional computational qubit are necessary to implement a parity-check-based CNOT gate \cite{parity,kerr}.  Five two-qubit gates (four additional computational qubits) are required to synthesize a cross-Kerr-based Fredkin gate \cite{kerr1}. Therefore, our schemes are superior to their synthesis-based counterparts, in terms of CNOT gates \cite{synthesis10,synthesis11}, and the cross-Kerr-based schemes \cite{Kerr4}. Moreover, large Kerr nonlinearity is a huge challenge for the current technology.  Additional computational qubits, necessary for parity-check-based and cross-Kerr-based universal gates \cite{kerr2}, are not required in our schemes. The evaluations of these two gates show that our schemes are feasible with current experimental technology.



\section*{Funding}

This work is supported by the National Natural Science Foundation of China under Grant No. 11604012, the Fundamental Research Funds for the Central Universities under Grant No. 230201506500024.

\end{document}